\documentclass[12pt]{article}
\usepackage{amssymb,amsmath,mathrsfs}
\begin{document}

\begin{titlepage}

                            \begin{center}
                            \vspace*{1cm}
\Large\bf{Extensive limit of a non-extensive entanglement entropy}\\

                            \vspace{2.5cm}

              \normalsize\sf    NIKOS \  \ KALOGEROPOULOS $^\ast$\\

                            \vspace{0.2cm}
                            
 \normalsize\sf Weill Cornell Medical College in Qatar\\
 Education City,  P.O.  Box 24144\\
 Doha, Qatar\\

                            \end{center}

                            \vspace{2.5cm}

                     \centerline{\normalsize\bf Abstract}
                     
                           \vspace{3mm}
                     
\normalsize\rm\setlength{\baselineskip}{18pt} 

\noindent An important calculation has been that of  the (von Neumann) 
entanglement entropy of the ground state of 1-dimensional lattice models at criticality and of their massive 
perturbations. This entropy turned out to be, generally, non-extensive. 
It was noticed, by contrast, that the Tsallis entropy of such systems can be extensive 
for a particular value of the non-extensive parameter. This value was expressed  
as function of the central charge of the underlying conformal field theory. 
We provide an answer to the question on why for central charge approaching infinity, 
extensivity of the von Neumann entropy of the ground state is restored. \\

                             \vfill

\noindent\sf  PACS: \  \  \  \  \  \ 03.65.Ud, \  03.67.Mn, \ 05.70.Jk, \   05.90.+m, \ 11.25.Hf   \\
\noindent\sf Keywords:  \  von Neumann entropy, \ Nonextensive entropy, \ Entanglement, \ Calabrese-Cardy formulae.   \\
                             
                             \vfill

\noindent\rule{8cm}{0.2mm}\\  
   \noindent $^\ast$ \footnotesize\rm E-mail: \ \  nik2011@qatar-med.cornell.edu\\

\end{titlepage}


                                                                                 \newpage

                                          \normalsize\rm\setlength{\baselineskip}{18pt}

                                                 \centerline{\large\sc 1.   \ Introduction}
                                                 
                                                                                 \vspace{5mm}

The entanglement entropy has attracted considerable attention in the literature during the last thirty years [1]-[4]. One reason is that it may 
be able to elucidate the statistical origin of black hole entropy [4] which has been an intriguing question since its discovery.  
As computations of such an entropy are usually quite difficult to perform in realistic systems, 
one is motivated to attempt them for simplified, toy-models, as is usually done in Physics. 
Such models when carefully chosen, are believed to encompass the essential features of  more realistic systems [1]-[12]. \\ 

Quite frequently, such paradigmatic systems are low-dimensional lattice models. Such models have the advantage of 
an already built-in spatial regulator, in the form of the lattice spacing, which allows us to avoid initially at least, thorny issues related 
to divergent contributions of the ultra-violet modes. Although such lattice models are obviously not Lorentz-invariant, 
their scale-invariant  behaviour at their critical point allows them to be analyzed by methods that can be formally seen as  
the Wick-rotated (Euclidean) analogues of the analysis performed on  Lorentz-invariant models.\\   

In a well-known work [9], the (entanglement) von Neumann entropy of the ground state of 1-dimensional lattice models, at criticality, was 
calculated.  The computation used a combination of a replica-method approach, functional integrals and operator product expansions in 
2-dimensional conformal field theories on the $n$-sheeted complex plane with branch cuts. 
These results were extended for systems away from criticality [9] by looking at them as massive deformations of the 
corresponding massless systems described by the conformal field theories. Despite a shortcoming of these calculations 
already pointed out by the authors themselves [9], it was found, among other things, that the von Neumann entropy of the ground state of 
such systems is, at appropriate limits, not extensive as a function of the system size and/or the correlation and thermal wavelengths.\\  

In an attempt to further examine the extensivity of this entanglement entropy, [13], [14] used the Tsallis entropy analogue 
of the von Neumann entropy for systems at criticaility, the latter of which relies on the Boltzmann/Gibbs/Shannon entropic form. 
The authors of [13], [14] used some of the intermediate results of [9], relied on the applicability of such results under analytic 
continuation to non-integer replica power and demanded extensivity of entropy in terms of the system size. They found [13]
that the Tsallis entropy can be extensive for a value of the non-extensive parameter \ $q$ \ given by     
\begin{equation}
   q_{ent} \ = \ \frac{\sqrt{9 + c^2} - 3}{c}
\end{equation}
where \ $c$ \ is the central charge of the underlying conformal field theory. A question that was asked then [13], [14] 
was to understand why 
\begin{equation}
    \lim_{c\rightarrow\infty} q_{ent} = 1
\end{equation} 
namely why does the usual von Neumann entropy of the ground state become extensive in terms of the system size, 
in the limit of an infinite central charge conformal field theory [15]. In the present work we address this question. 
We use mainly the conformal field theory (world-sheet) view, but also present a  comment about the string 
theoretical (sigma model) interpretation [16] of this conclusion.\\

                                                                  \vspace{8mm}


                                            \centerline{\large\sc 2. \  Some background and the question of interest}

                                                                    \vspace{5mm}

Consider [1] - [12] a system which is in a pure quantum state \ $|\Psi \rangle$. \ Let the corresponding density matrix be indicated by \
$\rho = |\Psi \rangle\langle\Psi |$. \ Divide the system in two parts labelled by \ $A$ \ and \ $B$. \ The reduced density matrix \ $\rho_A$ \
of \ $A$ \ is found by tracing \ $\rho$ \ over part \ $B$, \ namely 
\begin{equation} 
     \rho_A \ = \ tr_B \ \rho
\end{equation} 
The entanglement entropy is the von Neumann entropy  
\begin{equation}
    S_A [\rho_A ] \ = \ - \ tr_A \ \rho_A \log \rho_A
\end{equation}
One could interchange the roles of parts \ $A$ \ and \ $B$ \ in the above definitions and would  get \ $S_A = S_B$, \ 
therefore it makes sense to speak of the entanglement entropy of the system made up of parts \ $A$ \ and \ $B$ \ without further 
qualification. \\

It should be noted right away that the entanglement entropy is based on the classical Boltzmann/Gibbs/Shannon entropic form which for s 
discrete set of outcomes \ $i\in I$ \ with corresponding probabilities \ $p_i$ \  is given by
\begin{equation}
  S_{BGS} [\{ p_i \} ] \ = \  - \sum_{i\in I} \ p_i \log p_i
\end{equation}
More recently the Havrda-Charv\' {a}t [17], Dar\'{o}czy [18], Cressie-Read [19], [20], Tsallis [21], [22] 
(henceforth called just ``Tsallis" for brevity) entropy has attracted some 
attention [22]. This is  a one-parameter family of functionals labelled by \ $q\in\mathbb{R}$, \ 
called non-extensive or entropic parameter,  which is given by [21]
\begin{equation}
    S_q [\{ p_i \} ] \ = \ \frac{1}{q-1} \left( 1 - \sum_{i \in I} p_i^q \right)
\end{equation}
We use units where the  Boltzmann constant is set equal to one throughout this work. Clearly 
\begin{equation}
     \lim_{q\rightarrow 1} \  S_q \ = \ S_{BGS}
\end{equation}
An immediate extension of the von Neumann entropy to the Tsallis-inspired entanglement  entropy would be  
\begin{equation}
     S_q [\rho_A ] \ = \ \frac{1}{q-1} \left( 1 - tr_A \ \rho_A^q \right)
\end{equation}
It appears, by comparing (4) and (8), that calculating \ $S_q$ \ is technically harder than calculating \ $S_{BGS}$. \ This is probably true 
but only because the systems that the Tsallis entropy conjecturally describes are more ``complex" than the ones that \ $S_{BGS}$ \ 
describes [22]. Obviously one has to qualify the word ``complex"; we use it loosely to refer to systems that are out of equilibrium or that possess 
long-range spatial and temporal correlations, non-ergodic behaviour etc. The statistical behaviour of such systems is currently less 
understood than that of systems at equilibrium, with short-range correlations, exhibiting ergodic behaviour in phase space etc.\\

Suppose, however, that one is uncertain on whether to use \ $S_{BGS}$ \ of \ $S_q$ \ for a particular system. 
Then the calculation of either of these entropies may be equally difficult, depending on the method of calculation used. 
This was true of the cases discussed in [9], [10] and about the way they went about their derivations, which is very strongly 
reminiscent of the replica method extensively employed in statistical systems with disorder. The authors 
considered a one-dimensional system on a lattice [9] - [11], with lattice spacing \ $a$. \ 
Such a distance is a very convenient regulator for providing a natural cut-off for the high frequency (ultraviolet) modes that may make 
the entropy diverge. Let  \ $L$ \ be the spatial extent of the part of interest of the  lattice system. The rest of the lattice system can be finite, 
semi-infinite of infinite.  Time \ $t$ \ is assumed to be continuous. This, in turn, gave rise to thermal states upon its 
compactification to a circle of circumference \ $\beta = T^{-1}$ \ that can be formally seen as a Wick rotation to imaginary time \ $\tau = it$ \ 
providing the transition from indefinite (Lorentzian) to positive definite (Euclidean) signature of the $1+1$ dimensional space-time metric 
[9], [10]. Let \ $\mathcal{L}_E$ \ indicate the Lagrangian in the Euclidean space signature and \ $\mathcal{S}_E$ \ the corresponding action 
\begin{equation}
      \mathcal{S}_E \ =  \ \int_0^t \mathcal{L}_E \ dt 
\end{equation} 
Assume that the system has Hamiltonian  \ $\mathcal{H}$, \ corresponding Hamiltonian operator \ $\widehat{\mathcal{H}}$ \  and let \ 
$| \psi(x)\rangle $ \  indicate the common eigenvectors and \ $\psi(x)$ \ the corresponding eigenvalues of a complete set of commuting 
operators at \ $x$. \ Consider the partition function
\begin{equation} 
 \mathcal{Z} \ = \ tr \ e^{-\beta \widehat{\mathcal{H}}}
\end{equation}
The density matrix at inverse temperature \ $\beta$ \ is then formally given by the path-integral between initial \ $in$ \ and final \ $out$ \  
locations [9]
 \begin{equation}
     \rho \ = \ \frac{1}{\mathcal{Z}} \ \int [d\psi (x, \tau )] \ \prod_x \delta (\psi(x,0) - \psi_{in} (x_{in})) \ \prod_x \delta (\psi(x,\beta) - \psi_{out} (x_{out})) 
              \       \      e^{-S_E}
\end{equation}
As mentioned above, the time variable is compactified to a circle. One applies periodic boundary conditions on this circle to all the 
(bosonic) operators  involved. This identification, in effect, sews the two discrete edges of the rectangle
on which the path integral is defined transforming it into a cylinder of circumference \ $\beta$ \ in the imaginary time direction [9] - [11].  \\

Consider as subsystem \ $A$ \ the union of the \ $j$ \ disjoint intervals \ $A = (x_1, x_2) \cup \ldots \cup (x_{2j-1}, x_{2j})$. \ 
Then determining the 
reduced density matrix \ $\rho_A$ \ of \ $A$ \ amounts to performing the functional integral  (11) on the cylinder but the integration should not 
be over these $j$ intervals, as indicated by  (3). In effect, this means that the cylinder over which the path-integral is performed has to 
have cuts in the location of these intervals in the spatial direction at \ $\tau = 0$. \ Calabrese and Cardy [9], [10] 
proceed to compute the trace of integer \ $n\in\mathbb{N}$ \ powers of \ $\rho_A$. \ They notice that this amounts to performing the path-integral 
over an $n$-fold cover of the cylinder where periodic boundary conditions outside the cuts are imposed between consecutive sheets 
on the operators whose eigenvectors \ $|\psi \rangle$ \ appear in the path-integral. The operator content of the beginning of the first and the end 
of the last sheet should also be identified in this picture [9], [10]. 
Let the corresponding partition function be indicated by  \ $\mathcal{Z}_n(A)$. \ Then 
 \begin{equation} 
      tr \ \rho_A^n \ = \ \frac{\mathcal{Z}_n(A)}{\mathcal{Z}^n}
 \end{equation}
To determine the von Neumann (entanglement) entropy (4), one uses a formal analytic continuation to \ $n\in\mathbb{R}_+$ \ giving
\begin{equation}
     S_A [\rho_A ]  \ = \ - \lim_{n \rightarrow 1} \ \frac{\partial}{\partial n} \ tr \ \rho_A^n 
\end{equation}
which after using (12), gives 
\begin{equation}   
    S_A [\rho_A ] \ = \ - \lim_{n\rightarrow 1} \ \frac{\partial}{\partial n} \frac{\mathcal{Z}_n(A)}{\mathcal{Z}^n}
\end{equation}  
One immediately notices that to determine \ $S_A$ \ this method first calculates \ $tr \rho_A^n$ \ which is exactly what is required to  
compute the right-hand-side of (8), therefore the Tsallis entropy \ $S_q[\rho_A]$. \ Hence due to the method followed, computation of 
the Tsallis  entropy (8) is already part of the calculation of the von Neumann entropy (13).\\

Carrying out the path integrals in (11) is clearly a daunting task, even for lattice models on the line. To simplify is willing 
to consider systems at their critical point. Then, the scaling hypothesis assumes and the renormalization group approach proves, 
the scale invariance of physical quantities (properly re-scaled) at the critical point. 
Since the correlation length diverges at criticality, the only remaining system parameter through which extensivity can be expressed  
is that set by tthe length \ $L$ \ of the sub-system of interest. We assume, for simplicity, that the spatial extent of the whole lattice 
system is infinite, but substantially similar results are reached in the case of finite systems [9], [10]. 
At criticality, the lattice spacing \ $a$ \ becomes 
inconsequential, therefore the underlying structure can be  treated as a continuous space, namely as a Riemann surface 
(the complex plane in our case).
The underlying symmetry of the system is enhanced from that  of, at best, the semi-direct  product of a discrete 
translation at constant time and that of a continuous translation along the time axis, to that of the conformal invariance on the 
$n$-fold Riemannian surface with branch cuts along the indicated lines. Conformal invariance in two dimensions is an extremely strong 
condition because this is the only dimension in which the conformal group is infinite dimensional [15]. As result of  Noether's theorem,  
conformal invariance  provides an infinite number of conserved currents, 
which allow one to  even get explicit solutions in some cases, for correlation functions 
and any other quantities of physical interest [15].\\

 Due to the above, consider an interval of consecutive spins of length \ $L$ \ in the infinitely long 1-dimensional lattice system, which is assumed 
 to be  at its critical point at  \ $T=0$. \  Then the computation of the entanglement entropy for a conformal field theory of central charge \ $c$ \ 
 gives  [9], [10],  that 
 \begin{equation}
     tr \ \rho_A^n \ \sim \ \left( \frac{L}{a} \right)^{-\frac{c}{6} (n-\frac{1}{n})}
 \end{equation}
 Similar expressions have also been derived for the cases of finite temperature and for systems out of criticaility, the major difference being the 
 exact value of the denominator of the central charge in the exponent of (15). So, we will use (15) as the paradigmatic case since the other ones 
 are, admittedly highly non-trivial but still, variations of this one. Then [13] analytically continues this expression from \ $n\in\mathbb{N}$ \ to \ 
 $q_{ent}\in\mathbb{R}_+$ \ and asks for the value of such \ $q_{ent}$ \ so that the corresponding expression is extensive, namely 
 \ $S_{q_{ent}} \sim \ L$. \ This amounts to setting the exponent of (15) 
 
 \begin{equation}
     \frac{c}{6} \left( q_{ent} -  \frac{1}{q_{ent}} \right) \ = \ -1
 \end{equation}
 
                           \vspace{2mm}
 
 \noindent 
 which gives (1). Given (16), one observes that (2) holds. This states that the entanglement entropy of the ground state of the lattice system 
 becomes extensive for exactly the von Neumann functional form (4) for \ $c\rightarrow\infty$. \ So even though the Tsallis entropy (8) 
 is  extensive for the particular value of \ $q_{ent}$ \ given in (1) for any finite \ $c$, \ in the infinite central charge limit (2)
 one recovers the extensivity of the von Neumann/BGS functional form (4) . The question that was posed in [13], [14] was to explain this 
 restoration of extensivity of (4), expressed through (1) and (2), for \ $c\rightarrow\infty$.\\   
   
                                                                  \vspace{5mm}

                                                                

                                          \centerline{\large\sc 3. \ A proposed answer}

                                                                 \vspace{5mm}

To address this question, we will be far less general in our considerations. 
We will assume, for instance, that \ $c\rightarrow\infty$ \ by taking integer values. 
This  amounts to considering only the integer valued sub-sequences of all possible sequences of the central charge \ $c$ \ 
diverging to infinity. We do this, as some underlying features are simplified and this approach also allows for a ``stringy" (sigma model) 
interpretation of the final results.  The action (9) of a massless free boson, in a Wick-rotated (positive) signature metric, is 
\begin{equation} 
       \mathcal{S}_E \ = \ \frac{1}{4\pi \kappa } \ \int_{\Sigma} \ g^{\alpha\beta} (\partial_\alpha X) (\partial_\beta X) \ d\sigma
\end{equation}
where we are a bit general and indicate by \ $g_{\alpha\beta}$  \ the components of the metric tensor on the Riemann surface \ $\Sigma$ \ 
and where \ $d\sigma$ \ indicates the Riemannian area element of \ $\Sigma$. \  
The normalisation constant \  $\kappa$ \ is usually set to be one. In complex coordinates \ $z, \ \bar{z}$ \ this can be re-expressed as 
\begin{equation}
    \mathcal{S}_E \ = \ \frac{1}{4\pi \kappa} \ \int_\Sigma \ (\partial X)(\bar{\partial} X) \  dz \ d\bar{z}
\end{equation}
with \ $\partial \equiv \partial_z$ \ and \ $\bar{\partial} \equiv \partial_{\bar{z}}$. \ The (Hilbert) stress-energy tensor has non-vanishing 
components 
\begin{equation} 
  T_{zz} \ \sim \ \partial X \ \partial X, \hspace{15mm} T_{\bar{z}\bar{z}} \ \sim \ \bar{\partial} X \ \bar{\partial} X 
\end{equation}
with \ $T_{z\bar{z}} = 0$. \  It turns out that \ $T$ \ fails to be a primary (tensor)  field in the quantum theory, 
since it acquires a Schwinger term, expressing the breaking of  conformal invariance at the quantum level. 
This conformal invariance violation at the quantum level is quantified by 
the central charge \ $c$ \ of the conformal field theory [15].  Alternatively, on can look at the leading singularity  
in its operator product  expansion   
\begin{equation}
    T(z) T(w) \ \sim \ \frac{1/2}{(z-w)^4} + \ldots
\end{equation}
which gives that \ $c=1$ \ for the free boson, as is very well known [15]. Therefore the limit \ $c\rightarrow\infty$ \ effectively corresponds to 
putting an infinite number of free bosons on the Riemann surface. \\

The Hilbert space \ $\mathscr{H}$ \ of the free boson theory  is built by applying the Virasoro operators \ $L_{-k}, \  k >1$ \ 
to the vacuum,  which is a highest  weight state in a Verma module \ $\mathcal{V}$ \  (``representation") of the Virasoro algebra. 
Such a Verma module \ $\mathcal{V}$ \  is  labelled by the central charge \ $c$ \ of the conformal field theory and by the (conformal) 
dimension of its highest weight state \ $h$. \  Consider \ $\mathcal{N}$ \  free (non-interacting) bosons on \ $\Sigma$. \ 
We are interested in \ $\Sigma = \mathbb{C}$ \ with branch cuts. 
Then  \ $c_{tot}\rightarrow\infty$ \ amounts to \ $\mathcal{N}\rightarrow\infty$, \  since for a free boson \ $c=1$. \
Hence the Hilbert space of \ $\mathcal{N}$ \ free bosons on \ $\Sigma$ \ is the tensor product \ $\mathscr{H}^\mathcal{N}$. \ 
More generally one has 
\begin{equation}
     \mathscr{H} \ = \ \sum_{h, \bar{h}} \  \mathcal{V}_{c,h} \otimes \bar{\mathcal{V}}_{\bar{c}, \bar{h}} 
\end{equation}
where the summation extends to both the chiral/holomorphic and the anti-chiral/anti-holomorphic sectors. 
From (21), to compute the entanglement entropy of even  the ground state one needs an enumeration of all states of \ $\mathcal{V}$. \
It is straightforward to check that at each level \ $N$, \ 
the number of such states in the chiral/holomorphic sector, is given by the number of partitions \  $\mathcal{P}(N)$. \ 
The number of such partitions \ $\mathcal{P}(N)$ \  is given, for large \ $N$, \ by the Hardy-Ramanujan asymptotic formula
\begin{equation}
    \mathcal{P}(N) \ \sim \ \frac{1}{4\sqrt{3} \ N} \ \exp\left( \pi \ \sqrt{\frac{2N}{3}}\right)
\end{equation}
We observe that \ $\mathcal{P}(N)$ \  is (almost) an exponential function of  the level \  $N$. \ We can know consider \ $\mathcal{P}(N)$ \
for different values of \ $N$, \ which are succinctly, collectively encoded in the generating function   
\begin{equation}
      \mathcal{F}(s) \ = \ \prod_{M=1}^\infty \ \frac{1}{(1-s^M)} \ = \ \sum_{N=0}^\infty \ \mathcal{P}(N) \ s^N 
\end{equation}  
Combining the holomorphic and anti-holomorphic sectors gives the total generating (modular) function  
\begin{equation}
    \mathcal{F}(s, \bar{s}) \ = \ \prod_{M=1}^\infty \ \frac{1}{(1-s^M)(1-\bar{s}^M)}
\end{equation}
In more geometric terms, (23), (24) amount to  stating that the configuration or phase space volume
of the system, which in our case is \ $\mathscr{H}^\mathcal{N}$, \ grows exponentially as a function of its effective 
number \ $\mathcal{N}$ \ of degrees of freedom. This is  a feature of systems whose collective behaviour is given by the BGS entropic 
functionals (4), (5) as indicated by [23] and clarified in far greater generality in [24]. By contrast, the Tsalis entropy appears to be associated with 
systems whose configuration or phase space volume grows in a power-law (polynomial) manner with respect to the number of their effective 
degrees of freedom [23], [24].  Given this realisation,  the emergence of the BGS functional as a result of (1) and (2) by imposing extensivity 
of the entanglement entropy (8) as a function of the system size may not be all that surprising. \\

A complementary side of the above argument, which can be used to shed some additional light to it, 
 is  understanding why systems with finite \ $c_{tot}$, \ namely with finite \ $\mathcal{N}$, \ 
cannot be described by  the von Neumann  entropy (4), if extensivity  of this entropy is required. 
This can be understood by using (22). We observe that for finite \ $\mathcal{N}$ \ 
the maximum number of available states increases sub-exponentially with the level \ $N$. \ As a result,  there can not 
be exponential growth of the configuration or  phase space volume of the system as a function of \ $\mathcal{N}$. \  
Effectively, the system has already some built-in correlations that do not allow  an exponential growth rate 
of the system's phase space volume. Hence (4) is not suitable for describing the behaviour of the system and at the 
same time maintaining its extensivity [24]. This can also be seen by the form of the von Neumann entropy, which is  
 \begin{equation}
     S_A \ \sim \ \frac{c}{3} \log\frac{L}{a} + const.
 \end{equation}
 where \ $const$ \ stands for a non-universal constant.
The effective length of the system is actually \ $\log L$ \ rather than having the apparent value \ $L$ \ in this calculation. 
This  a result is stemming from the fact that not all of the allowed states of \ $\mathscr{H}$ \ actually contribute to the entropy. 
In other words, the number of effective degrees of freedom of the  system is less than what it superficially  appears to be. 
Only in the limit \ $\mathcal{N} \rightarrow \infty$ \  all such states can contribute to the entropy  
thus allowing the system to be described through (4). This argument  becomes a bit more concrete and transparent in the sigma 
model interpretation given in the last two paragraphs of this section. \\

We can ask if one can reach the same conclusion by looking at the finite temperature version of the above calculation [9] - [11]. 
As it was pointed out previously, the method of computation is similar to the zero temperature case and gives the following result [9]
for the von Neumann entropy of the system at temperature \ $\beta^{-1}$
\begin{equation}
     S_A \ \sim \ \frac{c}{3} \log \left( \frac{\beta}{\pi a} \sinh \frac{\pi L}{\beta} \right) + const'
\end{equation} 
We observe that (26) is extensive in the limit \ $L \gg \beta$. \ When, however, \ $L\ll \beta$ \ then we reach the same result (25), as in the 
zero-temperature case. This can be interpreted to mean that if the thermal wavelength, which for massless particles is proportional to \ 
$\beta$, is much larger than the sample size, then the von Neumann entropy is extensive. In other words, in the ultraviolet, where many energy 
modes of the system are accessible, (4) is extensive as pointed out in the previous paragraph. Here some care should be 
exercised though, for systems out of criticality, as in the ultraviolet the lattice spacing becomes progressively more important the more 
someone perturbs the underlying conformal field theory. But this does not affect the final result in the limit \ $\mathcal{N} \rightarrow\infty$. \ 
By contrast, in the infra-red fewer of the existing states of the system can be occupied, something that brings a severe restriction in the 
accessible volume of the phase space, which (as was previously pointed out) can only grow in a power-law manner [23], [24]. 
Then the Tsallis entropy  (8), but not the von Neumann entropy (4), 
is extensive. In a similar manner as above, if one considers an infinite number of free bosons, then there are enough (infinite) 
low-lying states that can be occupied even for \ $L\ll\beta$. \ As a result, there are no constraints to the  accessible phase space of the system  
whose volume can grow as fast as that of a gas of the free bosonic modes, namely exponentially as a function of \ $\mathcal{N},$ \  
a situation well-described by the BGS entropy.\\ 

From a sigma model viewpoint, the number of independent bosonic degrees of freedom on \ $\Sigma$ \ can interpreted as the dimension 
of space-time on which the Riemann surface \ $\Sigma$ \ is embedded.  
So, the calculation of the entanglement entropy really depends on the number of oscillatory string states, 
where the string is relativistic and embedded in \ $\mathbb{R}^\mathcal{N}$ \ space-time, for \ $\mathcal{N} \rightarrow\infty$. \ 
There are \ $\mathcal{N}-2$ \ directions of oscillation of the string in 
such a space-time, since the time-like and the longitudinal modes are excluded as unphysical/gauge dependent choices. 
The number of states \ $\mathcal{P}(N)$ \ at level \ $N$ \ that we seek, is the number of partitions of such \ $N$ \ allowing for 
\ $\mathcal{N}-2$ \ polarisation directions. This can easily be seen in the context of canonical quantisation of the oscillatory 
modes of the string. This number of partitions \ $\mathcal{P}(N)$ \ admits the asymptotic approximation 
\begin{equation}  
      \mathcal{P}_\mathcal{N} (N) \ \sim \ \left( \frac{\mathcal{N}-2}{24} \right)^\frac{\mathcal{N}-1}{4} \ N^{-\frac{\mathcal{N} + 1}{4}} \ 
                           \exp \left(2\pi \sqrt{\frac{N(\mathcal{N}-2)}{6}} \right) 
\end{equation}
We observe that \ $\mathcal{P}_\mathcal{N} (N)$ \  increases super-exponentially in terms of \ $\mathcal{N}$, \  so it becomes very fast the 
dominant term in (27). As a result of this super-exponential growth rate, the BGS entropy should be applicable in 
the \ $\mathcal{N} \rightarrow \infty$ \ limit, rather than the Tsallis entropy [24], as also noted before. Upon a more careful investigation,  
it can be noticed that  the underlying space-time is actually an orbifold having conical singularities induced by the branch cuts of the world-
sheet, rather than the Euclidean space \ $\mathbb{R}^\mathcal{N}$. \  Such identifications as the ones resulting in orbifolds with conical 
singularities   affect the number of accessible states, giving a number smaller than that of (27). However , given that such 
orbifold identifications can be traced back to non-free, properly discontinuous discrete group actions    
which  do not affect the asymptotic  behaviour of (27) as a function of \ $\mathcal{N}$. \ Therefore, the above conclusions 
remain unaltered by considering such space-time orbifolds instead of $\mathbb{R}^\mathcal{N}$.  \\       

Instead of relying on a set of free bosons, one could also take \ $c_{tot} \rightarrow\infty$ \ by considering fermions or even any of the 
parafermionic models whose central charges are expressed by the unitary minimal series [15]. We expect that the results would be largely 
the same as the distinction between bosons and fermions in two dimensions is not as sharp in higher dimensions. 
This allows for  the well-known bosonization procedure. What we would lose however by considering such anyonic 
conformal field theories would be the space-time  (sigma model) ``stringy" interpretation of the argument of the above paragraph, 
which we find intuitively appealing and useful in elucidating some of our conclusions.\\

                                                                \vspace{3mm}


                                              \centerline{\large\sc 4. \ Conclusion}

                                                                \vspace{5mm}

In this work, we addressed a question posed in [13], [14] about the recovery of extensivity of the von Neumann entanglement  entropy of the 
ground state of  one dimensional lattice systems at their critical point at zero and non-zero temperatures. We relied on the results of 
[9 ]-[ 11] which used methods of conformal field theories to arrive at the sought after formulae in terms of the central charge.
We also used in a substantial way the  results of [23], [24]  regarding the relation of Tsallis entropy to the rate of growth of 
the volume of phase space. We proposed an answer both in terms of the conformal field theory when it  is a set  of non-interacting bosons, 
something that also allows for a  ``stringy" (sigma model) interpretation of the results. \\ 

                                                                  \vspace{5mm}


                                                        \centerline{\large\sc References}
 
                                                                          \vspace{7mm}
 
 \noindent [1] \ R.D. Sorkin, \ in \ \emph{Tenth International Conference on General Relativity and Gravitation, 
                              \hspace*{6mm}         Contributed Papers, Vol. II}, \ p. 734 \ (1983). \ \ {\sf arXiv:}\\ 
 \noindent [2] \ L. Bombelli, R.K. Koul, J.-H. Lee, R.D. Sorkin, \ \emph{Phys. Rev. D} {\bf 34}, 373 \ (1986).\\
 \noindent [3] \ M. Srednicki, \ \emph{Rhys. Rev. Lett.} {\bf 71}, \ 666 \ (1993).\\ 
 \noindent [4] \ S.N. Solodukhin, \ \emph{Liv. Rev. Rel.} {\bf 14}, \ 8 \ (2011).\\ 
 \noindent [5] \ J. Cardy, I. Peschel, \ \emph{Nucl. Phys. B} {\bf 300}, \ 377 \ (1988).\\ 
\noindent [6]  \ C. Holzhey, F. Larsen, F. Wilczek, \ \emph{Nucl. Phys. B} {\bf 424}, \ 44 \ (1994).\\
\noindent [7] \  V. Korepin, \ \emph{Phys. Rev. Lett.} {\bf 92}, \ 096402 \ (2004).\\
\noindent [8] \ H. Casini, M. Huerta, \ \emph{Phys. Lett. B} {\bf 600}, \ 142 \ (2004).\\
\noindent [9]  \ P. Calabrese, J. Cardy, \ \emph{J. Stat. Mech.} {\bf 0406}, \ P06002 \ (2004).\\
\noindent [10]  \ P. Calabrese, J. Cardy, \ \emph{J. Stat. Mech.} {\bf 0504}, \ P04010 \ (2005).\\
\noindent [11] \ P. Calabrese, J. Cardy, \ \emph{Int. J. Quantum Inform} {\bf 04}, \ 429 \ (2006).\\ 
\noindent [12]  \ H. Casini, M. Huerta \ \emph{JHEP} {\bf  11},\  087 \ (2012).\\ 
\noindent [13] \ F. Caruso, C. Tsallis \ \emph{Phys. Rev. E} {\bf 78}, \ 021102 \ (2008).\\ 
\noindent [14] \ C. Tsallis, \ \emph{Eur. Phys. J. A} {\bf 40}, \ 257 \ (2009).\\ 
\noindent [15] \ P. Di Francesco, P. Mathieu, D. S\'{e}n\'{e}chal, \ \emph{Conformal Field Theory},  \ Springer-Verlag, \\
                                  \hspace*{8mm} New York (1997).\\  
\noindent [16] \ J. Polchinski, \ \emph{String Theory, Vol. 1}, \  Cambridge University Press, \ Cambridge, UK \ (1998).\\
\noindent [17] \ J. Havrda, F. Charv\'{a}t, \ \emph{Kybernetica} {\bf 3}, \ 30 \ (1967).\\
\noindent [18] \ Z. Dar\'{o}czy, \ \emph{Inf. Comp. / Inf. Contr.} {\bf 16}, \ 36 \ (1970).\\
\noindent [19] \ N.A. Cressie, T.R. Read, \ \emph{J. Roy. Stat. Soc. B} {\bf 46}, \ 440 \ (1984).\\
\noindent [20] \ T.R. Read, N.A. Cressie, \ \emph{Goodness of Fit Statistics for Discrete Multivariate Data}, \ Springer,\\ 
                                  \hspace*{8mm} New York \ (1988).\\  
\noindent [21] \ C. Tsallis, \ \emph{J. Stat. Phys.} {\bf 52}, \ 479 \ (1988).\\
\noindent [22] \ C. Tsallis, \ \emph{Introduction to Nonextensive Statistical Mechanics: Approaching a Complex \\
                                  \hspace*{8mm} World}, \ Springer, \  New York \ (2009).\\ 
\noindent [23] \ C. Tsallis, M. Gell-Mann, Y. Sato, \ \emph{Proc. Nat. Acad. Sci.} {\bf 102}, \ 15377 \ (2005).\\  
\noindent [24] \ R. Hanel, S. Thurner, \ \emph{Europhys. Lett.} {\bf 96}, \ 50003 \ (2011).\\  
 
 \end{document}